\documentclass[journal=jacsat,manuscript=communication]{achemso}
\usepackage{color}
\usepackage{epstopdf}
\usepackage{balance}
\usepackage[font=scriptsize]{caption}
\usepackage{fancyhdr}
\usepackage[flushleft]{threeparttable}
\usepackage{amsmath}
\usepackage{sectsty}
\usepackage{graphicx}

\let\oldmaketitle\maketitle
\let\maketitle\relax
\author{Hoang-Phuong Phan}
\affiliation[QMNC, Griffith University]
{Queensland Micro and Nanotechnology Centre, Griffith University, Queensland, Australia}
\email{hoangphuong.phan@griffithuni.edu.au}
\author{Tuan-Khoa Nguyen}
\affiliation[QMNC, Griffith University]
{Queensland Micro and Nanotechnology Centre, Griffith University, Queensland, Australia}
\author{Toan Dinh}
\affiliation[QMNC, Griffith University]
{Queensland Micro and Nanotechnology Centre, Griffith University, Queensland, Australia}
\author{Ina Ginnosuke}
\affiliation[University of Hyogo]
{Department of Mechanical Engineering, University of Hyogo, Hyogo, Japan}
\author{Atieh Ranjbar Kermany}
\affiliation[QMNC, Griffith University]
{Queensland Micro and Nanotechnology Centre, Griffith University, Queensland, Australia}
\author{Afzaal Qamar}
\affiliation[QMNC]
{Queensland Micro and Nanotechnology Centre, Griffith University, Queensland, Australia}
\author{Jisheng Han}
\affiliation[QMNC]
{Queensland Micro and Nanotechnology Centre, Griffith University, Queensland, Australia}
\author{Takahiro Namazu}
\affiliation[Aichi Institute of Technology]
{Department of Mechanical Engineering, Aichi Institute of Technology, Toyota, Japan}
\author{Maeda Ryutaro}
\affiliation[National Institute of Advanced Industrial Science and Technology (AIST)]
{National Institute of Advanced Industrial Science and Technology (AIST), Tsukuba, Ibaraki, Japan}
\author{Dzung Viet Dao}
\affiliation[QMNC, Griffith University]
{Queensland Micro and Nanotechnology Centre, Griffith University, Queensland, Australia}
\author{Nam-Trung Nguyen \vspace{1em}}
\affiliation[QMNC]
{Queensland Micro and Nanotechnology Centre, Griffith University, Queensland, Australia}

\title{Ultra-high strain in epitaxial silicon carbide nanostructures utilizing residual stress amplification \vspace{1em}}


\begin{document}

	
	
	
	


\twocolumn[
\oldmaketitle
\newpage
\vspace{2em}
\begin{@twocolumnfalse}
\begin{abstract}
Strain engineering has attracted great attention, particularly for epitaxial films grown on a different substrate. Residual strains of SiC have been widely employed to form ultra-high frequency and high Q factor resonators. However, to date the highest residual strain of SiC was reported to be limited to approximately 0.6\%. Large strains induced into SiC could lead to several interesting physical phenomena, as well as significant improvement of resonant frequencies. We report an unprecedented nano strain-amplifier structure with an unltra-high residual strain up to 8\% utilizing the natural residual stress between epitaxial 3C SiC and Si. In addition, the applied strain can be tuned by changing the dimensions of the amplifier structure. The possibility of introducing such a controllable and ultra-high strain will open the door to investigating the physics of SiC in large strain regimes, and the development of ultra sensitive mechanical sensors. 

\end{abstract}
\end{@twocolumnfalse}
]
\newpage

Strain engineering plays an important role in electronic devices \cite{strain1,strain2}. For instance, in complementary metal oxide semiconductor (CMOS) technology, scaling down the device to sub 100 nm regime encounters several challenges due to the degradation of carrier mobility \cite{strain_mobi1,strain_mobi2}. To enhance the performance of silicon CMOS with shrinking size, the introduction of strain into Si channel has been a key concept to compensate the decrease in carrier mobility \cite{strain_frequency1,strain_frequency2,strain_frequency3}. In addition, micro/nano electromechanical systems (MEMS/NEMS) have also benefited from strain engineering where the residual strain was applied between epilayers to enhance the resonant frequency ($f$) and the quality ($Q$) factor following Euler-Bernoulli theory \cite{Euler-Bernouli}. 

Silicon carbide, a wide band gap material, has showed its high potential in MEMS/NEMS applications \cite{SiC1,SiC2}. The large band gap makes SiC an excellent candidate for high temperature electronics applications where Si has limitation. The large Young’s modulus of above 300 GPa offers SiC resonators higher resonant frequencies in comparison to Si based counterparts \cite{SiC_Young1,SiC_Young2}. Other physical properties such as the piezoresistive effect and the thermoresistive effect in SiC have also been employed to develop mechanical and thermal sensors for structural health monitoring (SHM) \cite{piezo1,piezo2,piezo3,piezo4,thermal1,thermal2,thermal3,thermal4}. Cubic SiC (3C-SiC) is the only polytype, among over 200 SiC polytypes, that can be grown (epitaxially) on Si substrate and thus is the most suitable polytype for MEMS/NEMS applications \cite{SiC1,3C-SiC}. 
This allows the reduction of SiC/Si wafer cost as well as makes it compatible with the conventional micromachining process that developed for Si. The epitaxial growth of SiC on Si further results in a large residual stress within the SiC film due to the lattice and thermal mismatches between the SiC film and the Si substrate \cite{SiC_frequen1}. This residual stress can be potentially applied to enhance the $f\times Q$ of the resonators. Resonators with high frequencies up to several GHz and $f\times Q$ factors up to $10^{11}$  have been developed through the application of residual stress \cite{SiC_frequen1,SiC_frequen2,SiC_frequen3}. 

In most previous studies, the maximum strain induced into SiC ranged from 0.1\% to 0.6\% \cite{SiC_residual1,SiC_residual2,SiC_residual3,SiC_residual4}. It should be noted that the larger strain, the higher resonant frequency can be achieved. In addition, large strains up to 10\% could also lead to several significant changes in electrical/optical properties, which have been observed in several materials \cite{largestrain1,largestrain2}. A large stress could also cause unintended properties such as wafer bow and cracks, making the realization of ultra-high strain in SiC an challenging issue.
Therefore, to make extremely high strain in SiC possible, it is important to locally amplify strain in a specific area, while keeping the strain in the whole film in a relatively small regime.

We propose here a novel approach to induce extremely high strain into SiC by using a nano strain-amplifier. For the proof of concept, we demonstrate the feasibility of introducing a strain of approximately 8\% into a SiC nano-spring structures, which is to the best of our knowledge at least one order of magnitude larger than that reported in the literature. The possibility of inducing such a high strain will pave the path for experimental investigation of the physics of SiC in high strain regimes as well as for the development of highly sensitive SiC sensors. Our methodology should also be applicable for not only SiC but other materials including graphene and transition metal dichalcogenides (TMD), where high strain regimes are of significant interest.

\begin{figure*}[t!]
	\centering
	\includegraphics[width=6.6in]{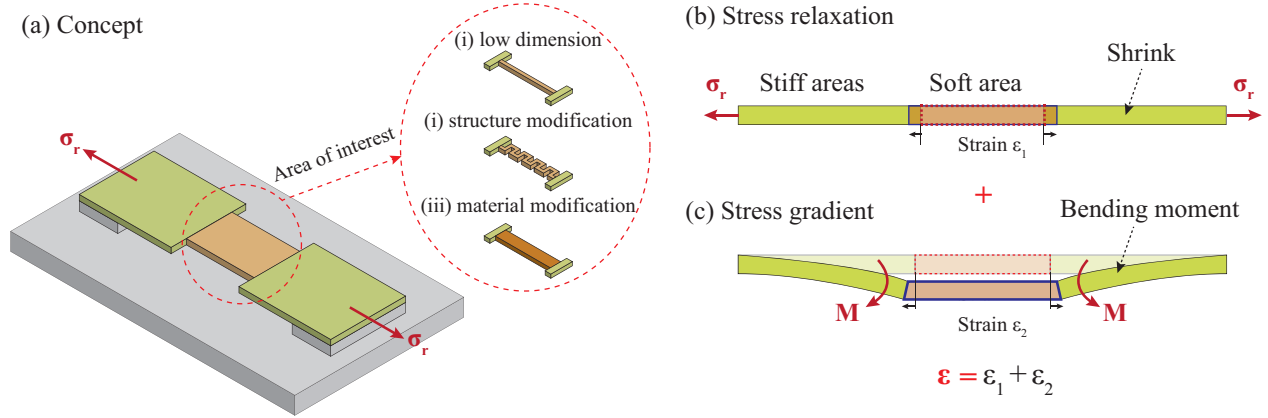}
	\caption{(a) Concept of devices to induce large strain into interested area; (b)(c) Principle of the nano strain-amplifier based on planar residual stress and stress gradient (side-view sketch).}
	\label{fig:concept}
\end{figure*}

Figure 1(a) illustrates the basic concept to induce large strain into SiC devices. The nano strain-amplifier consists of a SiC micro bridge which is released from a Si substrate, and a modified area which is located at the centre of the bridge. In addition, the modified area is designed to exhibit a smaller stiffness in comparison to the remaining areas of the SiC bridge. To soften this modified area, the following techniques can be utilized. First, the width of the middle area can be shrunk down to nanometer scale so that the ratio of the stiffness between the nanowire and micro-frame is significantly diminished. Second, the middle section can be softened by modifying its structure; as such a nano-spring can offer much larger elongation than that of a nano rod with the same dimension. Furthermore, by changing material from single-crystalline to amorphous, it is also possible to tune the stiffness of a specific region \cite{mater_modification1,mater_modification2}.  In the subsequent experiment, we selected the structure modification approach to form the soft area, as it can provide large elongation which eases the strain observation using scanning electron microscopy (SEM).

The key factor of our structure is that the low stiffness of the modified nano-area allows it to follow the deformation of the serially connected micro structures. As 3C-SiC has a smaller lattice size than Si, the SiC film typically undergoes a considerable tensile stress when being epitaxially grown on the Si substrate. When a SiC bridge is released, its lactic constant tends to return to its original size, which subsequently shrinks the size of the bridge. Because this phenomenon is more dominant in the SiC micro-areas, the nano-are will be stressed following the motion of these micro-area, Fig. \ref{fig:concept}(b).    
Furthermore, in a epitaxial film, large stresses typically distribute near the interface between top layer and the substrate. The residual stress gradually decreases with increasing film thickness or even relax for sufficiently thick films. This variation of the residual stress will cause a stress gradient along the thickness dimension, resulting in a bending moment once the film is suspended from the substrate. Therefore, under the residual bending moment, the micro-frames will deflect vertically, leading to an elongation in the softened nano-area, Fig. \ref{fig:concept}(c). Combining the two phenomena of in-plane shrinking and out-of-plane deflection, an ultra-high strain is expected to induce into the SiC nano structures. 

\begin{figure*}[t!]
	\centering
	\includegraphics[width=6.6in]{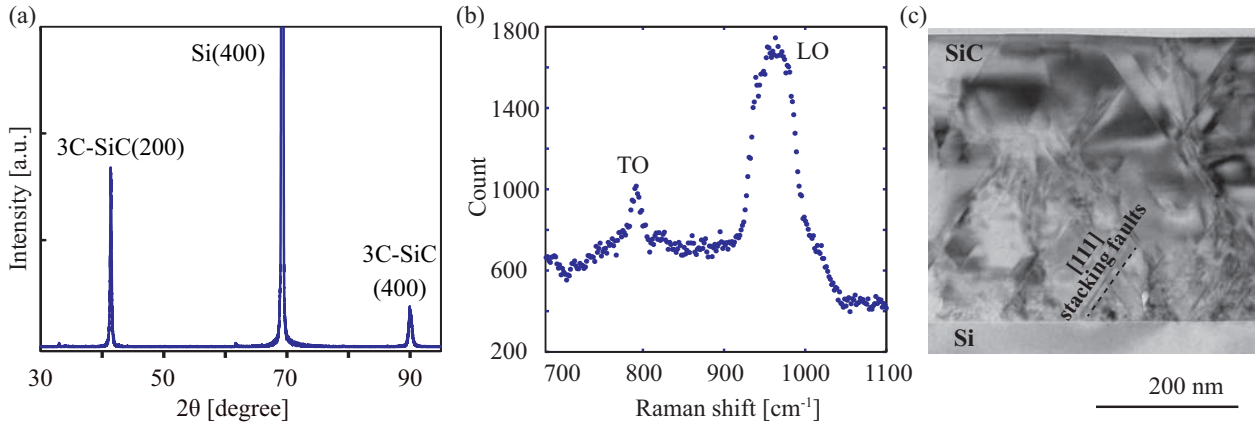}
	\caption{Properties of the SiC films. (a) X-Ray Diffraction; (b) Raman spectroscopy; (c) Transmission electron microscopy. }
	\label{fig:filmproperties}
\end{figure*}

The SiC film was prepared on a Si (100) wafer using low pressure chemical vapor deposition (CVD) where silane (SiH$_4$) and propylene (C$_3$H$_6$) were employed as the precursors \cite{SiC_growth}. As aforementioned, SiC and Si has significant lattice and thermal expansion mismatch of $\sim$20\% and $\sim$8\% respectively, the grown temperature was kept at 1000$^\circ$C utilizing alternating supply epitaxy (ASE) approach. X-Ray Diffraction (XRD) method was utilized to investigate the crystallinity of the SiC film. Figure \ref{fig:filmproperties}(a) shows diffraction peaks at 35.6$^\circ$ and 90$^\circ$, corresponding to 3C-SiC(200) and 3C-SiC (400) orientations. In addition, besides these two peaks, only a peak at 69.1$^\circ$ was observed, corresponding to Si(400). This result indicated that single crystalline (100) 3C-SiC was successfully grown on a Si substrate. The thickness of the films was found to be 294 nm using NANOMETRICS Nanospec/AFT 210.

The Raman scattering measurement was performed using \texttrademark Renishaw inVia Raman microscope 514 nm, Fig .\ref{fig:filmproperties}(b). The peak observed at 794 cm$^{-1}$ corresponds to the transverse optical (TO) vibration of 3C-SiC. Our TO peak was relatively smaller than that of free-stress 3C-SiC which was reported to be 795.9 cm$^{-1}$. Therefore, the left shift of the TO peak indicates the existence of a tensile stress in the epitaxial 3C-SiC film. The average tensile strain ($\varepsilon_r$) in the SiC film can be quantified using the coefficient reported by Rohmfeld \emph{et al.} \cite{SiC_residual1}:
\begin{equation}
\frac{\omega_{TO}}{\text{cm}^{-1}}= (795.9\pm 0.1)-(1125\pm 20)\varepsilon_r
\end{equation}       
Consequently, the average strain induced into the as-deposited SiC film is approximately 0.16\% to 0.2\%, which is in the same range with other epitaxial 3C-SiC film reported previously. Once, the SiC bridge is suspended, the residual stress is released and SiC tends to return to its free-stress state. Furthermore, as the SiC micro frames are much harder than the nano-spring (or other nano structures) located at the centre, the shrink of these frames are much more dominant. Assuming that the micro frames completely shrink down to their strain-free state, the strain induced into the nano-spring ($\varepsilon_{ns}$) can be approximated using the following equation:
\begin{equation}
\varepsilon_{ns} = \frac{L_f}{L_{ns}}\varepsilon_r
\label{eq:strain_compress}
\end{equation}  
where $L_f$ and $L_{ns}$ are the length of the frame and the nano-spring, respectively.

\begin{figure}[b!]
	\centering
	\includegraphics[width=3in]{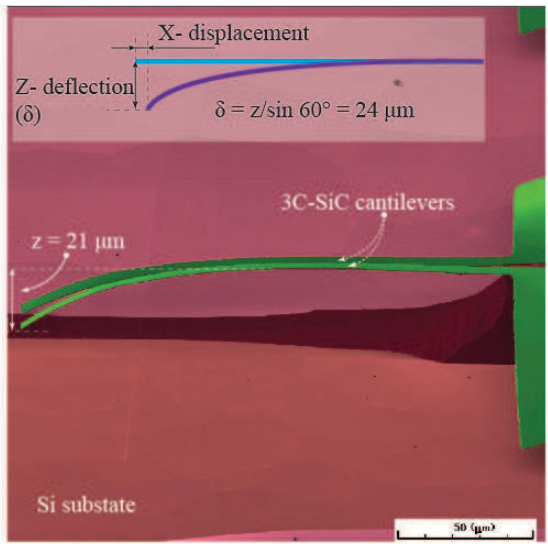}
	\caption{Observation of the out-of-plane deflection in a released 3C-SiC cantilever due to residual stress gradient (fall color).}
	\label{fig:cantilever_deflection}
\end{figure}

Additionally, the quality of the SiC film was characterized using transmission electron microscopy (TEM). The TEM image shown in Fig .\ref{fig:filmproperties}(c) indicates a high density of defects near the SiC/Si interface. Most defects were found to be the stacking faults in [111] crystallography orientation due to the lattice mismatch between SiC and Si. Quality of the SiC film increases evidently with increasing film thickness, exhibiting less crystal defect on the top layers. This observation also indicates a stress gradient through out the film. As such in our previous report, a large stress was measured at the bottom layers of 3C-SiC (100) films, while the stress decreased or relaxed at layers near the top surface \cite{strain_QMNC}. The bending moment exerted on SiC can be estimated from the Young's modulus of SiC and the deflection of free standing SiC cantilevers. Accordingly, the deflection ($\delta$) of a suspended SiC cantilever is given by:
\begin{equation}
\delta = \frac{Ml^2}{2EI}
\label{eq:deflect}
\end{equation}  
where $M$ is the bending moment; and $l$, $E$ and $I$ are the length, Young's modulus, and moment of inertia of the SiC cantilever. As a result, the bending moment caused by stress gradient is $M =2EI\delta/l^2 $. Figure \ref{fig:cantilever_deflection} shows a large out-of-plane deflection of a SiC cantilever due to residual stress gradient observed in a Focused Ion Beam (FIB) equipment with a tilt angle of 60$^\circ$. From the deflection of 24 $\mu$m, the dimension of the cantilever of 200$\mu$m $\times$ 5 $\mu$m $\times$ 0.3 $\mu$m, and the SiC Young's modulus of 350 GPa, the bending moment is estimated to be 5 $\mu$N.$\mu$m. Consequently, this relatively large bending moment excreted on the micro frames in nano strain-amplifier will cause a large vertical displacement, and thus results in a large tensile strain in the bridging nano-spring.    
 
\begin{figure}[b!]
	\centering
	\includegraphics[width=3.3in]{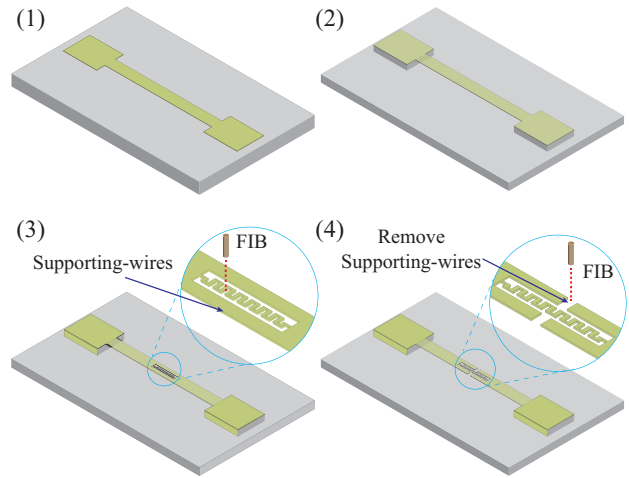}
	\caption{Fabrication process of the nano strain amplifier.}
	\label{fig:fab}
\end{figure}

Figure \ref{fig:fab} shows the fabrication process of the proposed devices. For the proof of concept we used a spring structure since it offers a lower spring constant in comparison to nanowires with the same width and thickness. The use of long SiC springs also allows a direct measurement of strain using a typical SEM. 

\begin{figure*}[t!]
	\centering
	\includegraphics[width=6.6in]{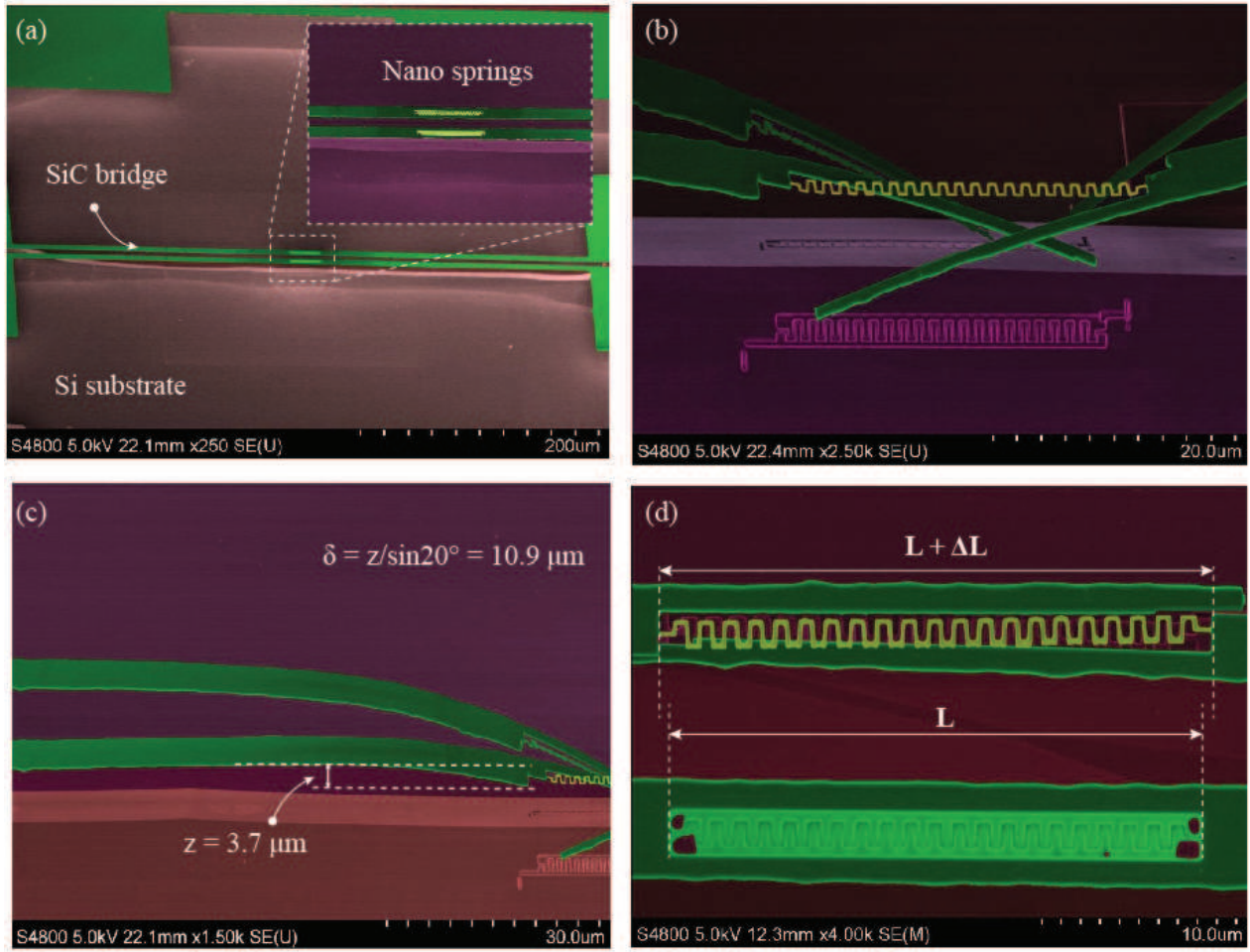}
	\caption{Colorized SEM images of SiC nano-springs. (a) Suspended SiC bridge with nano-spring fabricated at the center (prior to removing the supporting wires); (b) SiC nano-springs after cutting the supporting wires;  (c) Deflection of micro frame (or cantilever) on each side; (d) The elongation of the nano-spring observed from the top view.}
	\label{fig:photograph}
\end{figure*}
Initially, silicon carbide micro structures were patterned using inductive coupled plasma etching (HCl + O$_2$) with an etching rate of approximately 100 nm/min (step 1) \cite{SiCfabrication}. The SiC bridges were then released from the substrate by under-etching Si using TMAH at 90$^\circ$C for approximately 1 hour (step 2). The width of the bridges was fixed at 5 $\mu$m, while their length was varied from 100 $\mu$m to 500 $\mu$m. Consequently, SiC nano-springs and supporting wires were then formed at the middle of the bridge using focused ion beam (Ga$^+$) (step 3) \cite{SiC_FIB}. The length of the spring was approximately 27.71 $\mu$m, and the cross-sectional area was approximately 294 nm $\times$ 300 nm. Finally, the supporting wires was removed by FIB (step 3), releasing the nano-spring.   

Figure \ref{fig:photograph}(a) shows the SEM image of suspended SiC bridges prior to releasing the nano-spring. Evidently, the bridges were in almost flat shape, indicating a uniform strain contributing along its longitudinal direction. Figure \ref{fig:photograph}(b) presents the SEM image of the nano-spring after removing the supporting wires. The micro cantilevers on each side of the bridge were deflected vertically, leading to a large displacement in the nano-spring. Furthermore, the cantilevers were bent downward, indicating that a large compressive stress distributed at the bottom layer with respect to that of the top layers. This result is in good agreement with TEM observation mentioned above. Interestingly, because the two cantilevers were deflected at almost the same degree, the nano-spring remained in the lateral plane. Consequently the length of the nano-spring can be accurately measured from the top view SEM image, as shown in Figure \ref{fig:photograph}(d).
\begin{figure}[b!]
	\centering
	\includegraphics[width=3.3in]{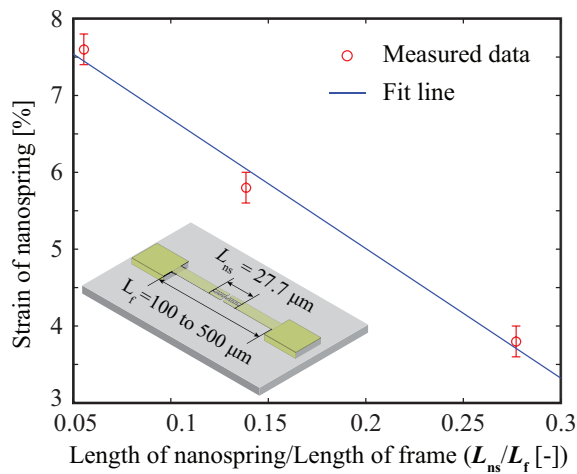}
	\caption{Strain in SiC nano-springs with different $L_{ns}/L_{f}$ ratios. For the same $L_{ns}$, increasing the length of the SiC bridge $L_{f}$ leads to an increase in the strain of the nano-spring.}
	\label{fig:strainvslength}
\end{figure}

Figure \ref{fig:strainvslength} plots the strain in SiC nano-springs against the ratio between the length of the nano-spring to that of the suspended bridge ($L_{ns}/L_f$). A strain of 7.6\% in the SiC nano-spring was observed when $L{ns}/L_f$) was 0.055, which is to the best of our knowledge the largest strain reported in SiC so far. Additionally, increasing the $L_{ns}/Ls$ results in a decrease in the applied strain, which is in solid agreement with Eq. \ref{eq:strain_compress} and Eq. \ref{eq:deflect}. More importantly, the dependence of the strain of the nano-spring on $L_{ns}/L_f$ indicates that ultra-high strain is controllable by changing the device dimensions. This could be applied to tune the resonant frequency in SiC nano resonators.

Another interesting property in our nano-strain amplifier is that the strain was only concentrated in the locally formed nano-spring, while the micro frame (i.e. the SiC cantilever on each side of the bridge) can be almost freely deformed releasing the residual stress. It means that the strain can be amplified at interested area, while a small strain is maintained at other area. Therefore, the proposed structure can be applied to locally induce extremely high strain into 2D materials such as graphene or transition metal dichalcogenide, where the Van der Walls adhesion force between them and the substrate (e.g. Si, PDMS, glass) typically cannot withstand 1\% strain \cite{2Dmater1,2Dmater2,2Dmater3,2Dmater4}.   

In conclusion, this work presents a novel mechanical approach to apply extremely large strain to SiC utilizing a nano strain-amplifier and the residual stress. We successfully demonstrated an extremely high strain up to nearly 8\% in a SiC nano-spring structure. The proposed platform shows its potential for further investigations into the physics of SiC nano structures in high strain regimes, along with the development of high frequency, high Q factor, and ultra-sensitive SiC mechanical sensors.


\vspace{1em}
\section*{Acknowledgment}
This work was performed in part at the Queensland node of the Australian National Fabrication Facility, a company established under the National Collaborative Research Infrastructure Strategy to provide nano and micro-fabrication facilities for Australia's researchers. 

\balance


\end{document}